\documentclass[journal,9pt]{IEEEtran}

\usepackage{amsmath}
\usepackage{cite}
\usepackage{graphicx}
\usepackage[caption=false,font=footnotesize]{subfig}
\usepackage{footnote}

\begin{document}

\title{Secure Optical Networks Based on Quantum Key Distribution and Weakly Trusted Repeaters}

\author{David Elkouss, Jesus Martinez-Mateo, Alex Ciurana, Vicente Martin
\thanks{The authors are with the research group on Quantum Information and Computation at Universidad Polit\'{e}cnica de Madrid, Spain.}
}

\maketitle

\begin{abstract}
In this paper we explore how recent technologies can improve the security of optical networks. In particular, we study how to use quantum key distribution (QKD) in common optical network infrastructures and propose a method to overcome its distance limitations. QKD is the first technology offering information theoretic secret-key distribution that relies only on the fundamental principles of quantum physics. Point-to-point QKD devices have reached a mature industrial state; however, these devices are severely limited in distance, since signals at the quantum level (e.g. single photons) are highly affected by the losses in the communication channel and intermediate devices. To overcome this limitation, intermediate nodes (i.e. repeaters) are used. Both, quantum-regime and trusted, classical, repeaters have been proposed in the QKD literature, but only the latter can be implemented in practice. As a novelty, we propose here a new QKD network model based on the use of not fully trusted intermediate nodes, referred as \textit{weakly trusted repeaters}. This approach forces the attacker to simultaneously break several paths to get access to the exchanged key, thus improving significantly the security of the network. We formalize the model using network codes and provide real scenarios that allow users to exchange secure keys over metropolitan optical networks using only passive components. Moreover, the theoretical framework allows to extend these scenarios not only to accommodate more complex trust constraints, but also to consider robustness and resiliency constraints on the network.
\end{abstract}


\IEEEpeerreviewmaketitle

\section{Introduction}

Optical network design has evolved over time to meet different challenges: high bandwidth, deployment flexibility, multiuser requirements, etc. Other characteristics were considered secondary at the time and added later, as their need or convenience became more pressing. Security has been one of these secondary requirements. It was usually taken for granted in optical networks due to the technical difficulty of spying the signal carried by an optical fiber~\cite{Senior_08}. However, technological advances and transformations in the structure of the network changed the panorama and this is no longer true~\cite{Shaneman_04, Fok_11}. On one hand, sensitive detectors and small transponders are nowadays able to perform, at a fraction of the cost and size, the operations that previously needed rack-sized equipment. On the other, certain network architectures simply does not lend itself to security~\cite{Gutierrez_07, Kazovsky_11}. As an example, downstream signals in a gigabit passive optical network (GPON) arrive to all users and are only dismissed by a well behaved optical network terminator (ONT). Nothing prevents them to actually record the signals.

Classically, there are several ways to provide security in a network. The most used one is just to add a cryptographic layer ---independent of the optical network layer--- that ciphers all the communications in the network. In a typical setting, a session key is exchanged in some way, for example using asymmetric (public-key) encryption and its underlying infrastructure (e.g. RSA and certification authorities) or by having exchanged previously a physical storage media with a pool of keys. Then, the exchanged key is used for a given maximum amount of time or of ciphered data size. When the pool is exhausted, a new one must be exchanged. In high security settings, no single channel is considered safe enough and a mixture of several methods (physical and RSA, and at different times using different paths) are used. While these are well-known techniques, they also have their drawbacks. For instance, the security of RSA is still an unproven assumption. Although its exponentially difficult nature has still to be really challenged\footnote{Except for a quantum computer that, in principle, is able to break the RSA. Whether such a computer will be built is still a matter of debate.}, the continuous growth in power of the algorithms and computers make for a constant revision of the recommended key size. What once was considered secure during the age of the earth, was actually broken in 17 years \cite{Atkins_95}. The recommendations for security have steadily grown from a few hundred bits to 2048 bits~\cite{RSA} length or even close to 15 kbits for certain operations that require a level of security equal to symmetric-key algorithms \cite{Barker_06}. These are even bigger when long term security ($\approx 20$ years) is required. Hence, if a technology is able to produce a continuous supply of high quality symmetric keys all over a network, it would be a welcome addition to its capabilities. In some cases as valuable as bandwidth itself. In this regard, the purpose of this paper is to explore a recent technology, namely quantum key distribution (QKD), to provide security in an optical network and, in particular, to propose some means to overcome its main limitation in distance through the application of ideas from network coding.

Quantum key distribution \cite{Gisin_02} allows two legitimate parties to generate a secret key even in the presence of an eavesdropper. The key is known only to the parties, since the information leaked to an eavesdropper can be bounded as tightly as they want. Hence, QKD is a technology able to distribute information-theoretic secure keys. The measurement of a quantum system in an unknown state modifies it (except if the same basis is used for measurement and coding---state preparation), thus allowing to detect the signals suspected to have been read by an eavesdropper. No computational complexity assumptions are needed. Just the laws of physics and, of course, common assumptions in cryptographic scenarios (e.g. the eavesdropper cannot control the devices at his will). A QKD protocol requires a quantum channel and an authenticated classical channel. The quantum channel is a communication channel supporting the transmission of quantum signals that are typically encoded as qubits: two states quantum systems such as the horizontal/vertical polarization states of a single photon. An optical fiber performs very well as a qubit carrier, hence a purely optical network is a very good candidate to implement a quantum network, i.e. a network based on the exchange of quantum signals.

QKD technology is delicate, since it must deal with signals at the quantum level. However, it has reached the point where devices able to work unattended during months are commercially available~\cite{Stucki_11, Jouguet_12}. As in any media, photons suffer an exponential attenuation while propagating through the fiber. In conventional communications, this is usually solved by using an amplifier, but in quantum communications this cannot be done, as amplifying a signal is just a measurement made to clone it, something forbidden at the quantum level by the ``no cloning'' principle: if a signal is unknown (i.e. we do not know for sure whether the qubit codes ``1'' or ``0''---the single photon is in a horizontal or vertical state of polarization, for example) the copy cannot be made exact \cite{Wootters_82}. A measurement made to obtain information results in an increase of the error rate in the signal, something that allows the detection of an eavesdropper.

Optical fiber is at the core of today's data networks because of its capability to support the bandwidth demand, relatively cheap manufacture, easy deployment and long life. The bulk of the communications is carried by optical fiber. Optical networks are pervasive nowadays, from long haul to fiber-to-the-home. Its bandwidth capacity has put much pressure on the electronics in order to keep pace with the raw transmission rate. Bandwidth, increase in reliability and low power consumption have made preferable to keep as much of the network in the purely optical domain. Therefore, most of the networks being deployed today are passive optical networks (PON). These can work under many different schemes, like the aforementioned GPON or WDM-PON\footnote{Wavelength Division Multiplexing-PON.}, but all of them have the characteristic that no active elements perturb the optical path---at least under typical metropolitan area distances. Thus, it is possible to create a direct, uninterrupted, optical path among two nodes in the network that is also able to support a quantum channel for QKD.

This allows to seamlessly integrate QKD in optical networks, where a quantum channel can be established between any two points in the network. In this regard, a flurry of activity has started in the field \cite{Townsend_94, Toliver_03, Lancho_09, Maeda_09, Chapuran_09, BingQi_10, Kitayama_11}; nevertheless, the distance limitation is still a problem. Current distance records are around 260 km for a link with experimental superconducting detectors \cite{Wang_12}. This means the capability to tolerate around 50 dB of losses, while commercial implementations are still not capable of going beyond 20 dB. If we take into account the insertion losses of typical optical components (e.g. filters, multiplexers, splitters, etc.), this barely allows to span a single access network\footnote{Although the new generation of QKD devices, able to withstand approximately 30 dB, is about to make their debut beyond the laboratories, reach is still a problem. Even assuming perfect detectors, emitters working at several GHz and new low losses fiber, there is no possibility of having any reasonable key rate beyond $\approx 500$ km.}. To alleviate these problems, specific networks devised for QKD have been proposed~\cite{Elliot_02, Peev_09, Sasaki_11, Kitayama_11, Stucki_11}. However, these are not really practical since, on one hand, the cost of using and deploying special infrastructures in populated environments, like in cities, is prohibitively high and, on the other, it does not solve the fundamental problem of limited reach.

Only quantum repeaters~\cite{Briegel_98} would allow to extend the reach without limits, but these are far beyond any practical technology today. Beyond quantum repeaters, we are forced to use trusted repeaters to overcome these losses. A trusted nodes network \cite{Alleaume_09} links two places through a series of shorter links that create a secret key. The key produced in the first link is relayed to the destination by encoding it with the keys created in the intermediate ones. Therefore, the key is known by all intermediate nodes, making the key secure only as long as all these are trusted. If a spy is successful attacking one, all the key is known. Since this is the only practical possibility, all network testbeds deployed up to date (e.g.~\cite{Elliot_02, Peev_09, Sasaki_11, Kitayama_11, Stucki_11}) make use of them in their design. The typical trusted node has a complex design \cite{Peev_09}, it includes several QKD devices (one to complete the pair needed in each link) and a computer to do all the associated key management, etc. which makes it even harder to certify~\cite{NIST_01, CommonCriteria} to any security level. The reliance on all these intermediate nodes makes trusted repeater networks expensive and not acceptable by many users.

We propose here a new approach to alleviate the reliance condition on trusted repeaters and apply it to the case of optical networks using standard components. Our work is based on the new paradigm provided by network coding \cite{Ahlswede_00}. The introduction of network coding by Ahlswede was a complete revolution in network theory: simple flow processing by the nodes allows to improve different scenarios in terms of throughput, needed resources and security. Here we use the idea of network codes to reduce the dependence on the trusted repeaters. We introduce conditional trust structures that guarantee that, as long as there is no cooperation between specified sets of \textit{weakly-trusted repeaters}, the distributed key remains fully secret. The name underscores the fact that some assumptions are no longer needed.

The paper is organized as follows. Section~\ref{sec:network-coding} reviews the basics of network coding and the wiretap network model that support our proposal. We have written the paper without assuming any previous knowledge about network coding, in consequence, this section is to a certain extent self-contained and should allow the reader to grasp the fundamental principles of network coding. In Section~\ref{sec:repeaters} we discuss the limitations of QKD and the need for intermediate repeaters. These repeaters are currently fully trusted, but using network coding principles we can impose a lighter trust constraint on the repeaters and achieve the same information theoretic security. In Section~\ref{sec:scenarios} we discuss the integration of weakly-trusted repeaters on passive optical networks and in Section V we describe some particular scenarios. Finally, we summarize the discussion and draw some future improvements in Section~\ref{sec:conclusions}.

\section{Network Coding}
\label{sec:network-coding}

Network coding is a paradigm where the intermediate nodes, instead of simply forwarding the incoming flows through the outgoing paths (according to some routing algorithm), distribute a function of the inputs through each outgoing path. It was shown in~\cite{Ahlswede_00} that linear combinations of the inputs suffice to maximize multicast transmissions. More generally, linear network codes allow to improve on several other aspects of common networks, in particular their security~\cite{Cai_11b}.

The application of network coding in optical networks has been widely studied in recent works. For instance, in~\cite{Manley_08, Kamal_10, Manley_10} the authors improve the performance, robustness and reliability of unicast and multicast optical networks, while \cite{Belzner_09, Miller_10, Fouli_11} focus on the new generation of passive optical networks commonly deployed in commercial infrastructures.

\subsection{Notation and Definitions}

Let us consider a network over a directed acyclic multigraph $\mathcal{G}$ defined by the tuple $(V,E)$, where $V$ is the set of vertices or nodes in the graph and $E$ the set of directed edges or links. We denote the adjacency in the network graph of the node $v$ to the edge $e$ using the notation $e \in \mathcal{A}(v)$. Two nodes in the graph, $v_{1},v_{2} \in V$, can communicate if there exist an edge $e$ such that $e \in \mathcal{A}(v_{1})$ and $e \in \mathcal{A}(v_{2})$. For convenience, we simplify the network model allowing every link to transmit a symbol taking values in the discrete finite alphabet $\mathcal{F}$. Note that by allowing multiple edges between two nodes we can generalize the model to links with different capacities.

The network serves a subset of nodes in the graph called source nodes $\mathcal{S}$. Every source $s \in \mathcal{S}$ generates a random message $M_{s}$ taking uniformly values from the discrete alphabet $\mathcal{M}_{s}$. We call $M$ the message jointly generated by all sources and consequently taking values in $\mathcal{M}$ the direct product of all $\mathcal M_{s}$.

\begin{figure}[!t]
\centering
\includegraphics[width=0.7\linewidth]{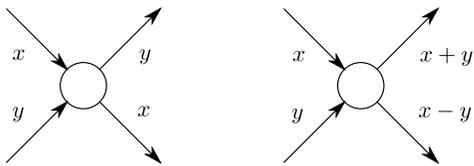}
\caption{Comparison of the forwarding paradigm with the network codes paradigm. The router on the left side of the figure only forwards the incoming packets, while the router on the right processes the inputs and transmits a linear combination of the inputs through every outgoing link.}
\label{fig:network-codes}
\end{figure}

In contrast to forward routing (see Fig.~\ref{fig:network-codes}), routers in the network coding paradigm are allowed to output a function of the incoming flows. If we restrict the functions to linear combinations of the inputs, we can easily deduce that they also represent linear combinations of the source messages. We associate every edge $e$ in the graph with $\phi_{e}$ a mapping from $\mathcal{M}$ to $\mathcal{F}$.

Finally, let $\mathcal{U}$ be the set of user nodes, a subset of nodes in the graph. A user $u \in \mathcal{U}$ aims to receive with no error $M_{u}$ the messages sent by $S_{u}$, a specific subset of $\mathcal{S}$. We will denote, abusing notation, by $Y_{e}$ and $Y_{u}$ the random symbol sent through the edge $e$ and the messages reaching the user $u$, respectively.

\subsection{Security}

\begin{figure}[!t]
\centering
\includegraphics[width=0.55\linewidth]{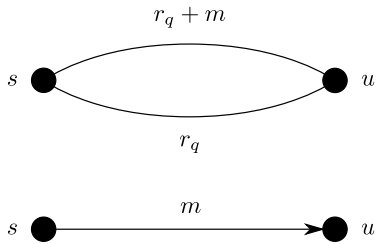}
\caption{A QKD link can be considered as composed by a private channel used for exchanging random secret keys $r_{q}$ and a public channel in which the key can help the source send a secret message $r_{q} + m$ to the user. It is equivalent to considering both as a single private link connecting the source and the user.}
\label{fig:qkd-link}
\end{figure}

Before we proceed to formalize the security framework, let us describe a network consisting of a single QKD link between two users (see Fig.~\ref{fig:qkd-link}). Logically, we can consider it to be composed by: (i) a private or secure link, in which a random key $r_{q}$ is exchanged, and (ii) a public channel in which a message $m$ is sent encrypted with a one time pad between $m$ and $r_{q}$. This is equivalent to having a private channel that a source $s$ can use to send $m$ to a user $u$. Therefore, we will consider every link in a QKD network to be a private link between its neighboring nodes. This restricts eavesdropping to the intermediate network nodes; only a curious router can gain access to network messages. This ability to extend the traditional security perimeter to also cover the communication channel between two QKD nodes is the consequence of the laws of quantum physics and is the key attribute of QKD. However, this property cannot be extended to classical repeaters. In essence, any classical repeater node in a chain of quantum links gets meaningful information \cite{Cai_11b}. In order to prevent the curious routers from accessing information, we can create extended source messages by adding randomness to the source messages. Formally, the messages are drawn from the direct product of the source alphabet $\mathcal{M}$, and a random key alphabet $\mathcal{K}$. The negative effect of the extended source messages is, however, a reduction in the achievable transmission rate, since the part devoted to the key transport increases the security but detracts from the information bandwidth.

Let us consider a set of $|\mathcal{W}|$ independent eavesdroppers. Every $w \in \mathcal{W}$ may receive the messages traversing a fixed collection of nodes, or eavesdropping pattern $B_{w}$, in order to recover a subset of the source message $M_{w}$. In consequence an eavesdropper has access to $Y_{B_{w}} = \{ Y_{e} : e \in \mathcal A(v), v \in B_{w} \}$, the messages traversing $B_{w}$. Note that the elements in $\mathcal{W}$, i.e. eavesdroppers, are elements of the power set of $V$ and in consequence potentially overlapping. We say that the intermediate nodes in the network graph are weakly trusted repeaters (WTR) to reflect the assumption that no further cooperation with the eavesdroppers is performed.

Following \cite{Cai_11b}, a network code is admissible over this \textit{eavesdrop network} model if every user node $u$ can recover $M_{u}$ and the information that every eavesdropper $w$ holds about $M_{w}$ does not reduce its entropy:

\begin{equation}
H(M_{w}|Y_{u}) = 0
\end{equation}

\noindent and $\forall w \in \mathcal{W}$:

\begin{equation}
H(M_{w}|Y_{B_{w}}) = H(M_{w})
\end{equation}

\noindent where $H(\cdot)$ and $H(\cdot | \cdot)$ are Shannon's entropy and Shannon's conditional entropy, respectively \cite{Cover_91}. These two conditions are called the secure and decodable conditions. Note that this is a generic definition, the special case in which there is one source, one user and the eavesdropper is interested in the whole message is just one of many possible configurations.

\subsection{Practical Scenarios} 
\label{sec:log-scenarios}

\begin{figure}[!t]
\centering
\includegraphics[width=0.7\linewidth]{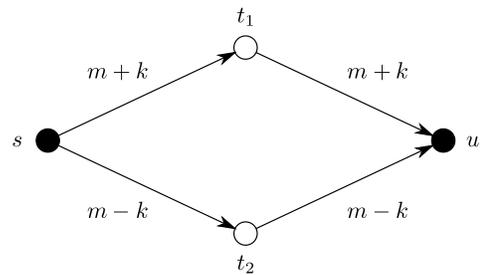}
\caption{In this network, the source $s$ wishes to send a message $m$ to the user $u$ (possibly a secret key) in the presence of $t_{1}$ and $t_{2}$, two intermediate nodes that eavesdrop their incoming and outgoing links. If $t_{1}$ and $t_{2}$ limit their eavesdropping activities to non cooperative eavesdropping they have no information about the source message $m \in \mathcal{M}$, where $k$ is a random message from the random alphabet $\mathcal{K}$.}
\label{fig:qkd-simple}
\end{figure}

A simple scenario that offers immediate gain for QKD using the network coding approach is shown in Fig.~\ref{fig:qkd-simple}. In this scenario two parties, the source $s$ and the user $u$, exchange a secure key relying in two intermediate nodes, $t_{1}$ and $t_{2}$ as depicted in Fig.~\ref{fig:qkd-simple}. The source generates a secret message $M$ and a secret key $K$ both taking values over the finite field $\textrm{GF}(3)$. If either $t_{1}$ or $t_{2}$ tries to get any information about $M$ it is easy to verify that

\begin{equation}
H(M|Y_{t_{1}}) = H(M|Y_{t_{2}}) = 0
\end{equation}

\noindent where $Y_{t_{1}}$ and $Y_{t_{2}}$ are the sets of extended messages traversing $t_{1}$ and $t_{2}$, respectively.

\begin{figure}[!t]
\centering
\includegraphics[width=0.7\linewidth]{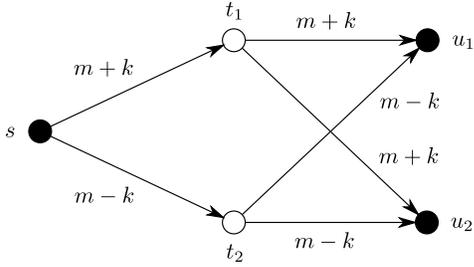}
\caption{In this figure the source $s$ distributes the same secret key to two different users $u_{1}$ and $u_{2}$.}
\label{fig:qkd-multi}
\end{figure}

The previous scenario can be used to enable multicast key distribution, as shown in Fig.~\ref{fig:qkd-multi}, without providing any further information to the intermediate routers: the extra links joining $t_{1}$ and $t_{2}$ with the second user $u_{2}$ replicate the links with $u_{1}$.

\begin{figure}[!t]
\centering
\includegraphics[width=0.85\linewidth]{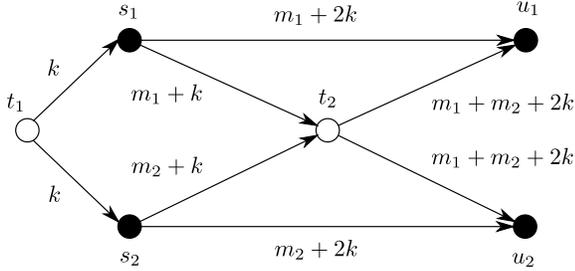}
\caption{This figure shows a network with two sources $s_{1}$ and $s_{2}$, and two users $u_{1}$ and $u_{2}$. The sources, $s_{1}$ and $s_{2}$, transmit $m_{1}$ and $m_{2}$ to the users, $u_{2}$ and $u_{1}$, respectively, while no information is leaked to the intermediate nodes or the remaining users.}
\label{fig:qkd-mac}
\end{figure}

Consider now the scenario depicted in Fig.~\ref{fig:qkd-mac}. In this scenario, proposed by Chan \textit{et al.} in~\cite{Chan_08}, four nodes ($s_{1}$, $s_{2}$, $u_{1}$ and $u_{2}$) exchange keys pairwise ($m_{1}$ between $s_{1}$ and $u_{2}$, and $m_{2}$ between $s_{2}$ and $u_{1}$) relying in one randomizing node $t_{1}$ and one centralized node $t_{2}$. In other words, $u_{1}$ and $u_{2}$ should be able to recover $m_{2}$ and $m_{1}$, respectively, but not $m_{1}$ and $m_{2}$. In effect, the users recover the desired message by adding the incoming flows and

\begin{equation}
H(M_{1}|Y_{u_{1}}) = H(M_{2}|Y_{u_{2}})=0
\end{equation}

It should be noted that $H(M|Y_{t_{1}}) = 0$, $H(M_{1}|Y_{t_{2}}) = H(M_{2}|Y_{t_{2}}) = 0$ but $H(M|Y_{t_{2}}) > 0$. That is, the network code is admissible if $t_{2}$ aims to recover either $M_{1}$ or $M_{2}$ but not both.

\subsection{Byzantine Adversaries}

In the analysis presented in the previous sections the adversaries have been described as passive eavesdroppers: they are assumed to be able to read the information stream, but not to modify it. This kind of bound on the eavesdropping power might be a concern because QKD, in absence of classical repeaters, provides their users with stronger security. However, this limitation on the eavesdropper is not actually a problem. In general, when dealing with the distribution of messages over a network, the security is studied against $t$-bounded adversaries, i.e. adversaries in control of at most $t$ nodes. These so called Byzantine adversaries are allowed to listen in their incoming links and output any message on their outgoing links.

A network with $t$-bounded adversaries is said to be secure if it provides the sources and users with perfect secrecy and perfect resiliency. Perfect secrecy is achieved if the adversaries get no meaningful information, in an information theoretic way, about the exchanged messages. Perfect resiliency means that $t$-bounded adversaries are unable to stop the sources from reliably transmitting the messages to the users. Dolev \textit{et al.}~\cite{Dolev_93} showed that if a network with one-way links has at least $3t+1$ node disjoint paths between a source and a user, the source can transmit messages with perfect secrecy and perfect resiliency (in the presence of $t$-bounded adversaries). In the network coding community Jaggi \textit{et al.}~\cite{Jaggi_08} showed, surprisingly, that if there are $3t+1$ node disjoint paths a source can transmit secure messages with rate $C-t$, where $C$ is the network capacity. That is, the only rate reduction from full capacity is $t$ and no extra penalty is paid.

Resiliency is a desirable property for QKD networks since a QKD link is in itself non resilient, i.e. QKD links are not protected from denial of service attacks. However, with current QKD networks dealing with small sets of nodes, the need for $3t+1$ node disjoint paths is a strong constraint: for instance, if the network is to be secured against any 3-bounded adversary, it has to provide 10 disjoint paths. Salvail \textit{et al.} propose in~\cite{Salvail_10} a weaker form of security: the network is secure if it provides with perfect secrecy and message authenticity. Message authenticity is achieved if either the message reaches the user uncorrupted or the user and the source are aware that it is corrupted. A message authentication schema is proposed such that in a network with $l$ disjoint paths, it can still provide message authenticity if there is at least one uncorrupted path, the network can provide message authenticity. Their schema is said to be unforgeable because no intermediate node, or set of nodes, outside from the nodes belonging to the uncorrupt path have any information about the exchanged messages. This same kind of restriction can be imposed to the network code scenarios; eavesdropping patterns can be defined such that the eavesdroppers can tap any node except for nodes belonging to one uncorrupted path. With these eavesdropping patterns the authentication mechanism can be applied with the same security.

\section{Quantum Key Distribution Networks}
\label{sec:repeaters}

QKD devices use qubits as their information carriers. These qubits are described mathematically as vectors in a two dimensional Hilbert space, while in practice, any two dimensional quantum system can be used to encode a qubit. When qubits are used in a quantum communication system, such as a QKD device, photons are the usual choice for the physical realization. Properties like polarization, phase or even time can be used to encode a qubit into a single photon, making them a flexible choice that are also reasonably easy to transmit, detect and manipulate: laser diodes, avalanche photo diodes, beam splitters, modulators, etc., are all the optical components needed. Photon qubits are transmitted through either free space (air) or optical fiber, being the latter the logical choice for communication networks. As any signal propagating through an absorbing medium, photons suffer an exponential attenuation when traveling through the fiber. This losses are harmful for the extremely sensitive QKD devices, which are made to transmit and manipulate single photons. Moreover, the interaction of the photon qubit with the environment is actually indistinguishable from an eavesdropper manipulation, thus rendering the signal unusable for cryptographic purposes. This includes any attempt to amplify the signal, which is basically an interaction with the qubit in order to know its state and copy it in many photons, something that cannot be done without introducing an error, the same error that is used to rule out the existence of an eavesdropper and that forms the basis of a QKD protocol. Under these conditions, a QKD device is limited to use fully transparent optical networks without active devices such as amplifiers or electro-optical converters and, even in this case, its maximum reach is limited.

\subsection{QKD Performance}

There are many factors that hinder the performance of a QKD system: far from ideal single-photon detectors or emitters are the two main ones; but even if they could be made perfect and working at a very high speed rate, practical limits of QKD systems would be ultimately set by the absorptions in the quantum channel. A system working at 10 GHz and with detectors 100\% efficient\footnote{Compared to the 5 MHz and 10\% quantum detector efficiency of the systems commercially available today} would have a maximum practical reach of 500 km using the best fiber available and working in 1550 nm, the most transparent window.

\begin{figure}[!t]
\centering
\includegraphics[width=\linewidth]{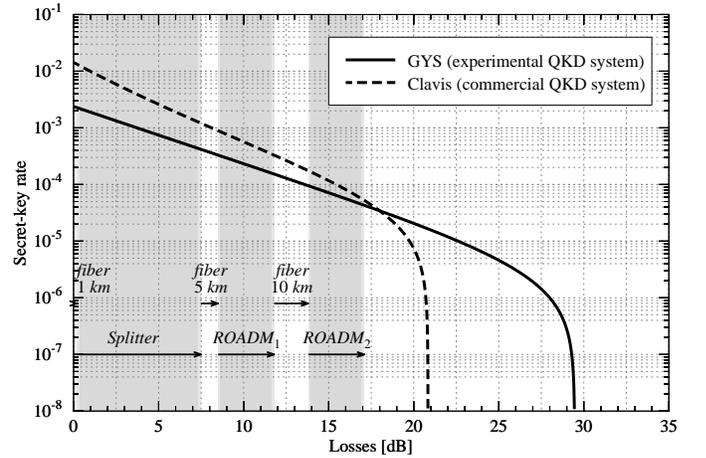}
\caption{Secret-key rate, in bits per qubit sent, of two different QKD systems, GYS and Clavis, using the BB84 protocol with decoy states as a function of the absorptions in the network. The ratio of secret key is calculated using the asymptotic approximation proposed in~\cite{Ma_05}. The parameters considered for the GYS system can be found in~\cite{Gobby_04}. In the Clavis system~\cite{idQuantique}, we use an absorption value of $0.25$ dB/km for the fiber (both devices are transmitting at the 1550 nm window). Losses due to network devices are depicted using a shadowed region. Note that the different regions are plotted in the order in which an hypothetical photon qubit would find them if it is connected through an access network homed into a passive metropolitan ring and crosses two core nodes before being dropped to its final destination.}
\label{fig:secret-key-rate}
\end{figure}

As an example of the practical QKD limits, we calculated the secret key rate using two different fiber based QKD systems, GYS~\cite{Gobby_04} and Clavis~\cite{idQuantique}, and present them in Fig.~\ref{fig:secret-key-rate}. GYS was selected because is a laboratory implementation with typical components and parameters that has been widely used in the literature as a benchmark, while Clavis is a QKD system already available in the market.  The latter is the development variant of the id Quantique Vectis system, used to provide commercial grade security based on QKD. The specific parameters for the systems are provided in the caption of Fig.~\ref{fig:secret-key-rate}.

Although it is typical to present the secret-key rate as a function of the distance in km, we show it as a function of the losses, which is the meaningful figure when using QKD in networks instead of a direct point to point link. This allows us to show (grayed out areas) the absorptions of common network components. This highlights the fact that in real optical networks they are the limiting factor. Fig.~\ref{fig:secret-key-rate} depicts the key exchange of two nodes that are located in different access networks of a metropolitan area network (its structure is detailed in Section~\ref{sec:scenarios}). The absorptions are presented in the same order in which a real qubit would find them.

Following the standard procedures for computing the secret key rate, the quantum bit error rate is roughly approximated as a function of the losses and the dark count rate which is assumed to be constant ($1.7 \times 10^{-6}$ in the GYS experiment and $2 \times 10^{-5}$ for the Clavis system). Since losses reduce the noise to signal ratio, the secret key rate in Fig.~\ref{fig:secret-key-rate} can also be regarded as a function of the quantum bit error rate.

The secret key rate shows the amount of secret key extracted per transmitted qubit. To calculate the secret key rate per second, it has to be multiplied by the emitter frequency ($5-10$ MHz for current commercial QKD systems). As expected, the secret key rate decreases exponentially with the losses up to a point where it goes to zero rather abruptly. This effect has nothing to do with the optical network, but with the way in which the secret key has to be extracted. When losses are higher, the signal to noise ratio is worse and the quantum bit error rate grows. Since in cryptography one has to deal with the worst possible case, all errors have to be attributed to an attacker. More quantum bit error rate means, then, that in the last step in a QKD communication, known as privacy amplification, a high number of bits have to be discarded to account for the presence of an hypothetical eavesdropper.

In any case, Fig.~\ref{fig:secret-key-rate} clearly shows that there is a maximum number of losses that can be tolerated. This is approximately 20 dB for the commercial system. This loss budget, that in a point to point link could amount to 80 km of standard optical fiber, is not enough to cross (without repeaters) a typical metropolitan area network. It should be noted that continuous variable QKD is an alternative technology for using in future telecom networks since it is intrinsically more adapted to WDM \cite{BingQi_10}, and although it has classically been considered highly limited in distance, recent experiments show that this technology is also capable of achieving the aforementioned 80 km \cite{Jouguet_12b}. New types of QKD devices could also be well suited for network integration \cite{Alba_12}. It is to be noted that tolerance to losses and high bit rates are complementary magnitudes, since high speed systems would be, in general, able to reach further away with a useful key rate.

\subsection{Trusted Repeaters}

Modern optical networks have embraced the full optical domain model, hence they are capable of transporting quantum signals. There have been many studies trying to integrate QKD in optical networks~\cite{Toliver_03, Lancho_09, Maeda_09, Chapuran_09, Kitayama_11}. However, the distance limitation still persists. As mentioned in the introduction, there are only two ways to overcome this. The best one is to build a repeater or amplifier in the quantum regime. This is a possibility not ruled out by theory and with good experimental progress \cite{Briegel_98} but one that, by all accounts, is still many years in the future. The other one is to build a classical repeater that is actually a measurement device that relays the keys produced in the first quantum link.

\begin{figure}[t]
\centering
\subfloat[Detail of the messages exchanged with a trusted repeater.]
{
\includegraphics[width=0.7\linewidth]{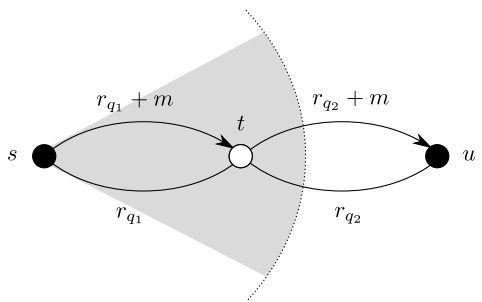}
\label{fig:qkd-simple-real}
}
\hfil 
\subfloat[Logical schema of a link extended with a trusted repeater.]
{
\includegraphics[width=0.7\linewidth]{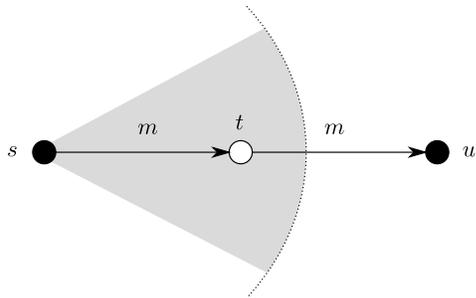}
\label{fig:qkd-simple-logic}
}
\caption{The reach can be extended by allowing trusted intermediate nodes to forward the key. The grey area marks the distance limitation of the QKD system with emitter source at $s$.}
\label{fig:qkd-simple-bothfigs}
\end{figure}

To extend the reach of QKD, an intermediate classical repeater can be used (see Fig.~\ref{fig:qkd-simple-real}). It is to be noted that a QKD link is composed of an emitter side and a receiver side. The emitter is typically made up of a laser diode while the receiver contains the single-photon detectors, which are based on avalanche photo diodes. Thus, a classical repeater, $t$, is composed of a receiver and an emitter (other pairing possibilities exist, but here we stick with the most simple one). The receiver links with the previous emitter in the chain and the emitter with the forthcoming receiver (in the figure, $s$ and $u$, respectively). The receiver is used to establish a quantum channel different from the quantum channel built up with the emitter. Both quantum channels create, independently, keys $r_{q_1}$ and $r_{q_2}$ with the same length. Then, through the public authenticated channel, a message $m$ is relayed securely computing a one-time pad with the QKD keys created before (the logical scheme of this process is depicted in Fig.~\ref{fig:qkd-simple-logic}). This process can continue with any number of intermediate links, but each repeater $t$ will know the message being transmitted. The secure message $m$ can be used later as a key to cipher future communications. Unfortunately, it only takes one corrupted node to completely spoil the security of the system. This full trust condition is not acceptable in many cases.

\subsection{Weakly Trusted Repeaters}

A message is information theoretically secure if the entropy of the message is not reduced with respect to a well-defined adversary and scenario. In QKD the eavesdropper is allowed to perform any attack allowed by quantum physics. However, the laboratories of the legitimate parties are assumed to be trusted and their devices well known. Recent studies show that it is possible to extract a non zero, though significantly smaller, secret key without making any assumption on the device characterization. This model is known as device independent QKD \cite{Masanes_11}. The secret key distilled by a device independent QKD protocol improves the security of a key distilled by a QKD protocol only in the sense that it reduces the set of hypothesis for information theoretic security.

If we use trusted repeaters to extend the reach of QKD the key is still information theoretically secure provided that the eavesdropper is limited to quantum attacks and the (well-characterized) laboratories and devices of Bob and Alice as well as all the intermediate repeaters are outside of Eve's control. However, the full disclosure implied by trusted repeaters might be too strong in many scenarios. In these situations, trusted repeaters are not a valid option. In contrast, we can only weakly trust the intermediate nodes and consider that some might try to recover information of the secret key or message, i.e. the setting discussed in Section~\ref{sec:network-coding}. At design time the set of tappable nodes and possible associations is established. The set can be delimited in a rather precise fashion, e.g. any set of $\ell$ nodes can be tapped, etc. Then, the users can search for a secure network code that fulfills their trust requirements. Nonetheless, the weakly trusted repeater paradigm does not increase the security with respect to trusted repeaters in a quantifiable way, what it offers is the possibility to modulate the number of intermediate repeaters that a user is willing to trust. It is in the same sense that device independent QKD improves on QKD that the weakly trusted repeaters paradigm improves on the security of the trusted repeaters scenario.

Explicit code constructions in the general wiretapping model is an open problem~\cite{Chan_08}. However, in the single source scenario, secure network code constructions are fairly well known~\cite{Cai_11b} and, in simple cases like the examples from Section~\ref{sec:network-coding}, they can be discovered by inspection. Hence, we can readily use this formalism to reduce the dependence on the intermediate nodes.

\section{Weakly Trusted Repeaters on Passive Optical Networks} 
\label{sec:scenarios}

\begin{figure}[t]
\centering
\includegraphics[width=0.7\linewidth]{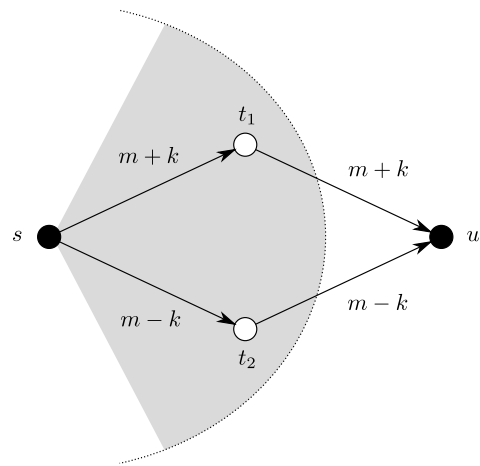} 
\caption{The scenario corresponds to an optical network with two disjoint paths that are used to overcome the QKD distance limitation. As in Fig.~\ref{fig:qkd-simple-bothfigs}, the grey sector marks the distance limitation of the QKD system with emitter source at $s$.}
\label{fig:qkd-simple-wtr}
\end{figure}

Assuming a network where nodes can be deployed at will, we can implement directly the first example from Section~\ref{sec:network-coding} (see Fig.~\ref{fig:qkd-simple-wtr}). The distance limitation of a QKD emitter at the source $s$ is denoted by a gray sector. This limitation can be overcome by introducing two disjoint paths between the source, $s$, and the user, $u$. In contrast to the trusted repeaters solution, different messages are sent through each path, such that here the key is not fully disclosed to any of the intermediate nodes.

Unfortunately, despite its benefits, practical optical networks do not have this deployment ease. Instead, nodes have to be arranged in specific structures to fit different geographical, cost, deployment, etc. constraints. The next scenario focuses on one of these kind of optical networks: metropolitan area networks.

As already mentioned in the introduction, optical networks are the preferred technology for commercial telecommunication networks. For instance, in metropolitan areas, optical fiber and passive technology is widely available, making possible to establish an optical path among two nodes. Here we focus on metropolitan area networks using only passive technology, since due to the moderate absorptions, optical paths do not need amplification. This means that they are not disrupted at the quantum level, hence being suitable for QKD.

\begin{figure}[!t]
\centering
\includegraphics[width=0.9\linewidth]{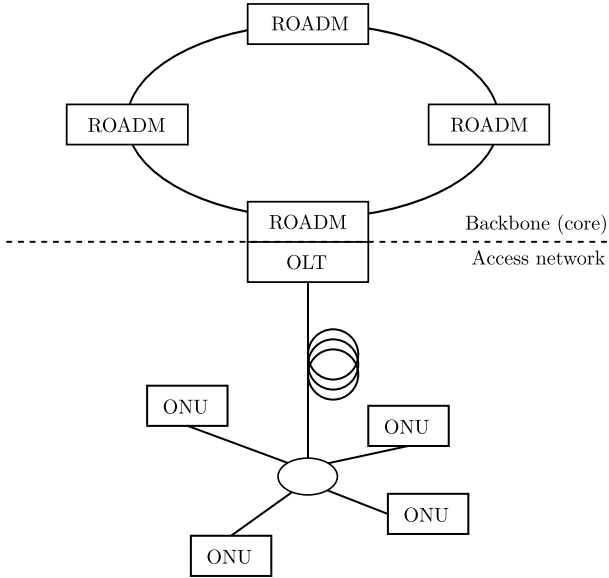}
\caption{Example of a typical metropolitan optical network. The core ring is depicted on top, while the lower part represents the point to multipoint access network.}
\label{fig:pon-1}
\end{figure}

Metropolitan area networks span from ten to several hundred kilometers. Fig.~\ref{fig:pon-1} shows the typical\footnote{Actual networks can be much more complicated, reflecting the competition over time among technologies and specific growth needs. We consider here only the typical model as representative enough.} design of such networks: a central ring-shaped backbone connected to peripheral access networks. The backbone is composed of multiple nodes, separated by tens of kilometers, that connect to the optical line termination (OLT) of the access networks. These nodes use different kinds of optical add-drop multiplexers (OADM), for instance reconfigurable OADM, to route signals to the correct access network depending on their wavelength. Starting at the OLT, the access networks follow a point to multipoint topology in order to serve nodes, known as optical network units (ONU), located in the same zone. The preferred choice for access networks is passive technology, thus the signal is distributed among ONU via passive devices like splitters o arrayed waveguide gratings (AWG). The typical distance between the backbone node and the ONU is a few kilometers~\cite{Ramaswami_09}. Using this structure, metropolitan area networks can accommodate many users via multiplexing techniques such as wavelength division or time division multiplexing (WDM and TDM respectively).

The use of QKD in optical network is further complicated if quantum and classical signals are multiplexed. Since at the quantum level we are dealing with single photon signals, any leak of the classical signals, typically 100 dB stronger, will spoil the quantum channel. Although there are studies demonstrating this coexistence (e.g. \cite{Lancho_09, Choi_10, Eraerds_10}), its effect basically amounts to increase the noise and, thus, reduce the maximum distance/rate achievable. Being the only purpose of this section to show the benefits of weakly trusted repeaters in optical networks, we prefer to keep the network in the quantum regime and avoid these problems. Thus we assume the network is being used only for QKD purposes. An authenticated classical channel is supposed to be available among any nodes.

To secure a metropolitan optical network, a quantum channel has to be created among any two nodes of the access networks. Therefore, first we connect an emitter to the end of an access network. Looking at Fig.~\ref{fig:secret-key-rate} and Table~\ref{tab:insertion-losses}, we realize that losses, due to fiber and network components, do not allow to directly plug the receiver in a different access network. Intermediate nodes are needed. Possible locations are the immediate backbone node or its closest neighbors. In the associated graph, this scheme generates a bipartite graph with emitters placed at the end of the access network and receivers at the backbone nodes. Each emitter has several outgoing links: to the receiver in its own backbone node and the neighboring ones. The type of QKD device selected for each node is not arbitrary. Receivers are more expensive and difficult to maintain due to the single-photon detectors, hence they are kept at the telco installations.

\begin{savenotes}
\begin{table}[!t]
\renewcommand{\arraystretch}{1.3}
\caption{Insertion Losses} \label{tab:insertion-losses}
\centering
\begin{tabular}{l l l}
\hline
Device & Operation wavelength & Insertion loss\footnote{These values belong to low-losses components (except for fiber, that we use the value for the installed in typical links) available in the market.} \\ \hline
Single-mode fiber & 1550 nm & $0.25$ dB/km \\
$1:2$ Splitter & 1260 -- 1610 nm & $3.5$ dB \\
$1:4$ Splitter & 1260 -- 1610 nm & $7$ dB \\
CWDM add-drop multiplexer & 1270 -- 1610 nm & $0.6$ dB \\
DWDM add-drop multiplexer & 1525 -- 1610 nm & $0.6$ dB \\
Bandpass filter & & $0.7$ dB \\
Circulator & 1530 -- 1565 nm & $0.5$ dB \\
Connectors & & $0.2$ dB/pair \\
AWG (40 channels) & 1525 -- 1610 nm & $3$ dB \\ \hline
\end{tabular}
\end{table}
\end{savenotes}

This shows that telecom networks not only suit QKD, but they are flexible enough to provide several alternative paths among two nodes (at least in metro areas). Therefore, the network coding approach described in Section~\ref{sec:network-coding} can be used. This implies that a commercial optical network with improved security and resilience, as compared to the traditional scheme, can be designed using QKD and WTR.

\section{Network Prototypes}
\label{sec:prototypes}

A simple scenario where two parties communicate through two intermediate nodes can be derived from the theoretical framework described in Section~\ref{sec:network-coding}. In this scenario the parties are able to exchange a secret-key under weak trust assumptions. This simplicity also facilitates the implementation of secure optical networks, like the ones described in the previous section. Here we propose several network prototypes to demonstrate that secure telecom PON can be implemented using QKD and WTR in metropolitan area networks (see Fig.~\ref{fig:pon-1}).

\begin{figure*}[!t]
\centering
\includegraphics[width=0.7\linewidth]{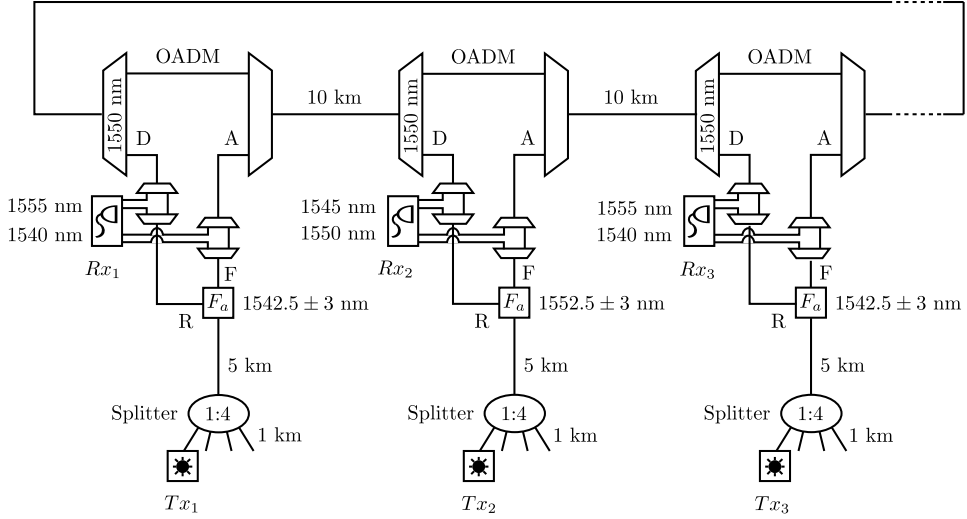}
\caption{Scheme of a QKD-PON under the network coding paradigm. Coarse WDM (CWDM) add-drop modules are used to provide a channel in the third transmission window, around 1550 nm, corresponding to the lowest absorption band. Dense WDM (DWDM) add-drop modules are used to route signals to the corresponding receiver. Finally, a band-pass filter $F_a$ is used to connect the access network with the backbone and route signals into the correct direction within the backbone.}
\label{fig:qkd-pon-1}
\end{figure*}

A first prototype is depicted in Fig.~\ref{fig:qkd-pon-1}. Emitters, labeled as $\mathrm{Tx}_n$, are connected to user nodes in an access network, and receivers, labeled as $\mathrm{Rx}_n$, are connected to backbone nodes. In a backbone node, several passive WDM components are included to route the quantum signals. The objective is to reach from an emitter at least two different receivers in order to have several paths for improved security. A coarse WDM OADM (CWDM) is used to route signals within a passband to the corresponding access network. Add and drop ports, $A$ and $D$ in the figure, are connected to a dense WDM OADM (DWDM). The function of this second multiplexer is to filter signals with the wavelengths associated to a receiver\footnote{This is one of many possible configurations, for instance, DWDM OADM could be replaced by a series of band-pass filters.}. In this wavelength addressing scheme, quantum signals can reach a receiver from two different directions. Both DWDM OADM are connected to a band-pass filter, $F_a$, using the reflected and filtered ports. The common port of the $F_a$ filter connects the backbone with the access network, thus routing signals from the access network to the correct destination within the backbone ring. In the access network several emitters are connected to the splitter. Fiber lengths are assumed to be within the typical distances for metropolitan area networks.

Even though this prototype works under the WDM paradigm, a wavelength multiplexer cannot be used because of the backbone CWDM OADM. Instead, an splitter is used, as in GPON. In this way, multiple emitters can communicate simultaneously with different receivers, because each receiver is addressed using a different wavelength. In particular, in this prototype each emitter can communicate with three receivers, all of them within the range of current commercial QKD systems (Fig.~\ref{fig:secret-key-rate}): the receiver in the immediate backbone node and both neighboring backbone nodes. For instance, in Fig.~\ref{fig:qkd-pon-1}, $\mathrm{Tx}_2$ is able to communicate with $\mathrm{Rx}_1$, $\mathrm{Rx}_2$ and $\mathrm{Rx}_3$. Since communications between farther nodes are unfeasible because of the absorptions, wavelengths for the filters $F_a$ and DWDM OADM can be used repeatedly all over the network. This reduces the number of required wavelengths and simplifies the network construction.

An example to illustrate the operation mode of this prototype is as follows. Assume that $\mathrm{Tx}_1$ and $\mathrm{Tx}_2$ cannot communicate directly because of the absorptions and then they want to exchange a key using weakly trusted repeaters, thus they need at least two intermediate nodes:

\begin{enumerate}

\item $\mathrm{Tx}_1$ transmits at 1540 nm. The signal is filtered by $F_a$ and it is dropped at the DWDM OADM, reaching then $\mathrm{Rx}_1$. The total loss budget, according to Table~\ref{tab:insertion-losses}, is $\approx 10.6$ dB.

\item $\mathrm{Tx}_1$ transmits at 1545 nm. The signal is filtered by $F_a$, passes the DWDM OADM, exits the backbone node, is dropped by the CWDM OADM and is dropped again by the DWDM OADM in front of $\mathrm{Rx}_2$. The total loss budget is $\approx 15.5$ dB.

\item $\mathrm{Tx}_2$ transmits at 1545 nm. The signal is reflected by $F_a$ and dropped by the DWDM OADM in order to reach $\mathrm{Rx}_2$. The total loss budget is $\approx 10.6$ dB.

\item $\mathrm{Tx}_2$ transmits at 1540 nm. The signal is reflected by $F_a$, it passes through the DWDM OADM, exits the backbone node, is dropped by the CWDM OADM and DWDM OADM, and then reaches $\mathrm{Rx}_1$. The total loss budget is $\approx 15.5$ dB.

\end{enumerate}

\begin{figure}[t]
\centering 
\subfloat[$\mathrm{Tx}_1$ to $\mathrm{Tx}_2$]
{ 
\includegraphics[width=0.7\linewidth]{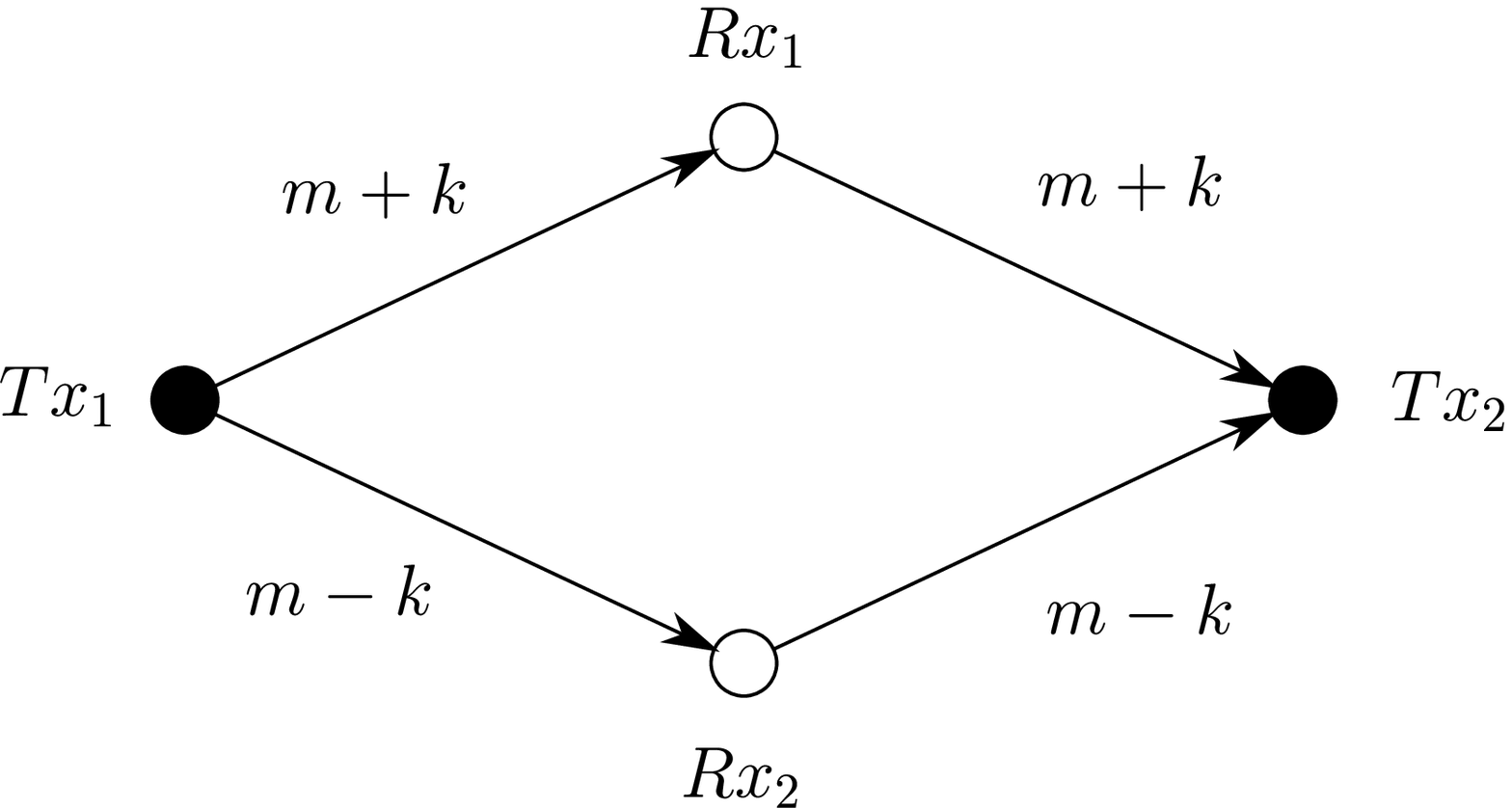}
\label{fig:qkd-scenario-nc-1}
} 
\hfil 
\subfloat[$\mathrm{Tx}_1$ to $\mathrm{Tx}_3$]
{ 
\includegraphics[width=0.95\linewidth]{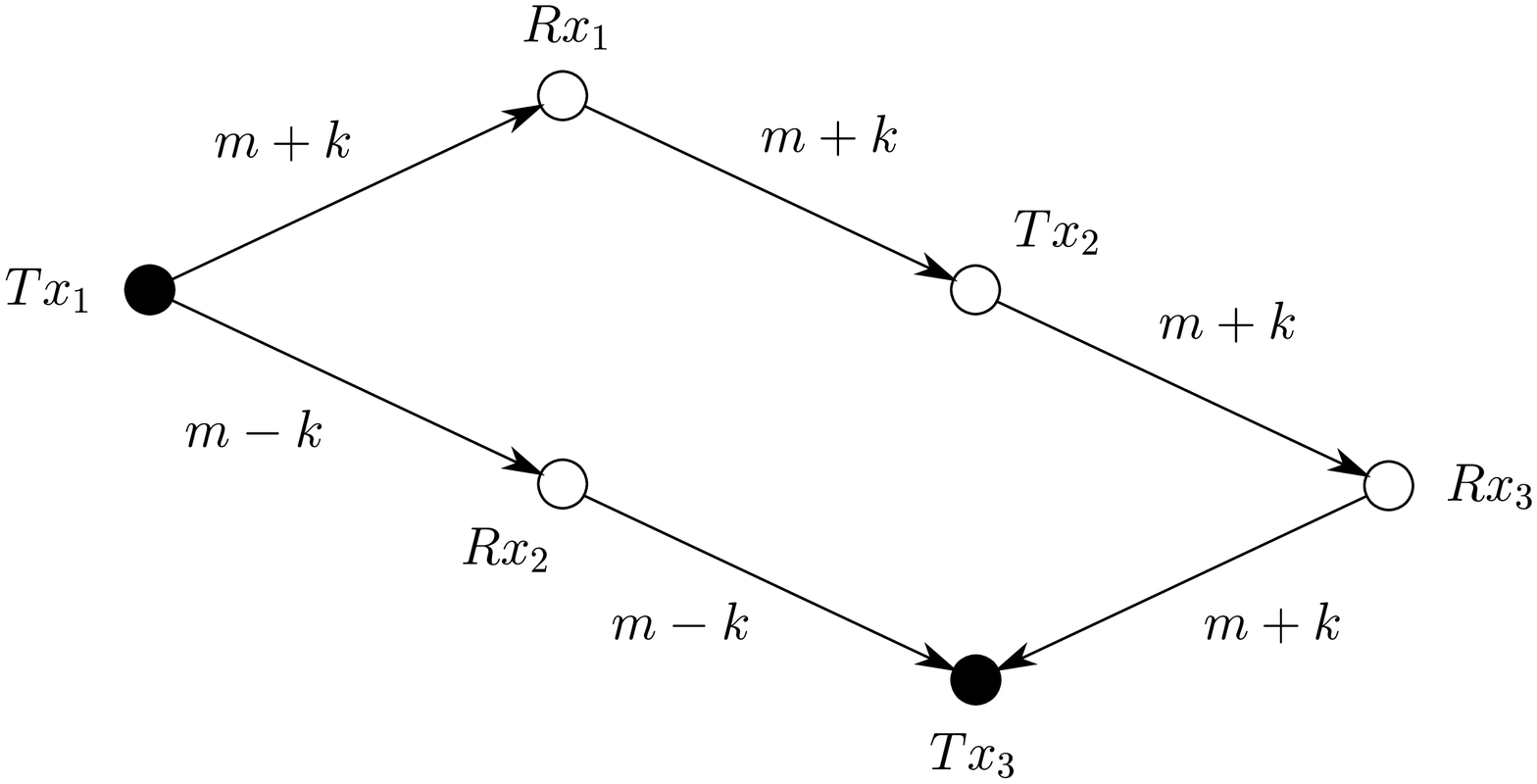}
\label{fig:qkd-scenario-nc-2}
} 
\caption{Diagrams for network coding purposes corresponding to key exchanges between nodes $\mathrm{Tx}_1$ and $\mathrm{Tx}_2$ (top) and between $\mathrm{Tx}_1$ and $\mathrm{Tx}_3$ using the optical network depicted in Fig.~\ref{fig:qkd-pon-1}.} 
\end{figure}

The logical diagram of the communication is presented in Fig.~\ref{fig:qkd-scenario-nc-1}. Note how it reflects the network coding structure of Fig.~\ref{fig:qkd-simple}. Using the secure keys exchanged through QKD, a secret message $m$ can be exchanged between $\mathrm{Tx}_1$ and $\mathrm{Tx}_2$ through authenticated classical channels. No information is disclosed to the intermediate (repeater) nodes: $\mathrm{Rx}_1$ and $\mathrm{Rx}_2$. They act as WTR and they gain no information about the message $m$ although they know either $m+k$ or $m-k$. This simple case can be modified to exchanges between farther nodes. For instance, Fig.~\ref{fig:qkd-scenario-nc-2} shows the case where $\mathrm{Tx}_1$ exchanges a key with $\mathrm{Tx}_3$. Nodes connected to the same splitter can be handled in a similar way.

\begin{figure}[!t]
\centering
\includegraphics[width=0.9\linewidth]{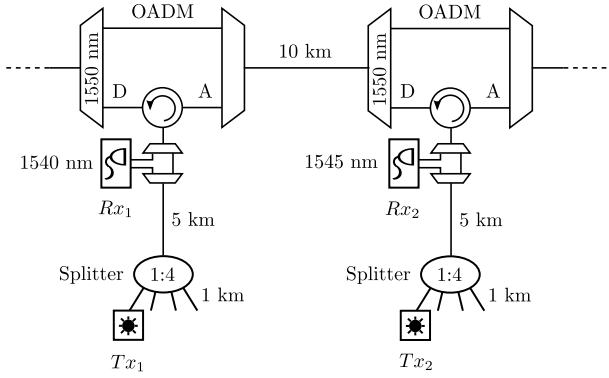}
\caption{Alternative scheme of a QKD-PON with circulators instead of band-pass filters. This reduces the number of components, but signals can travel only in the forward direction. This means that, in order to recover the connectivity with the previous node, accessible in the scheme of Fig.~\ref{fig:qkd-pon-1}, either the quantum signal has to travel the full ring in the forward direction till reaching the node before or a second ring running in parallel but in the backward direction must be used.}
\label{fig:qkd-pon-2}
\end{figure}

Fig.~\ref{fig:qkd-pon-2} shows a second network prototype. This has the advantage of requiring less components than the prototype depicted in Fig.~\ref{fig:qkd-pon-1}. The band-pass filter $F_a$ used to route signals within the backbone is replaced by a circulator, such that signals can now be transmitted in only one direction. In such a configuration, an emitter can reach only the receiver attached to its immediate backbone node and the first backbone node to the right. This limitation can be avoided in two ways: (i) assuming that the backbone network is a closed ring, such that any emitter can exchange keys with at least one forward receiver. Thus, two disjoint paths can be drawn using the full ring. (ii) Using an additional identical backbone network to transmit signals in the opposite direction. Note that in this prototype only one DWDM OADM is required, but this advantage disappears if we set up a second network to communicate in the opposite direction.

\begin{figure}[!t]
\centering
\includegraphics[width=0.8\linewidth]{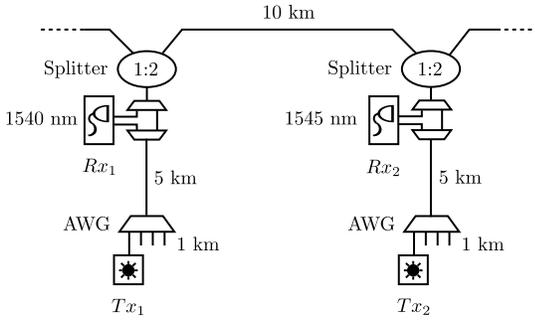}
\caption{Third prototype of a QKD-PON with splitters instead of circulators to solve the problem of only forward communications. Moreover, the number of OADM is reduced. These changes permit to use an AWG instead of an splitter, and thus increase the number of users at the access network. The disadvantage is that this scheme does not allow key exchanges beyond the nearest backbone node (e.g. $\mathrm{Tx}_1$ is not able to directly connect beyond $\mathrm{Rx}_2$). This is not an option with current commercial QKD systems, since their loss budget is as much 20 dB, but with the expected new generation, able to withstand 30 dB, this advantage ---except for the increase in throughput between nearest neighbor nodes--- would be lost. In this case, other schemes like the first two prototypes can be used.}
\label{fig:qkd-pon-3}
\end{figure}

A third network prototype is shown in Fig.~\ref{fig:qkd-pon-3}. This improves in terms of the number of users, resources and loss budget. Note that reducing the losses we are also increasing the final secret-key rate in a QKD system, thus, increasing the overall WTR transmission rate too. The circulator is replaced by an splitter, solving the problem of backward communications. CWDM OADM are removed since emitters and distant receivers cannot exchange secure keys because of the absorptions. This could be a disadvantage for the next generation of QKD systems which are able to withstand at least 30 dB. In this case, other schemes like the first two prototypes can be used to exchange keys between distant nodes. As a result, backbone nodes require only three components: splitters, OADM and receivers. Now, the splitter in the access network can also be replaced by an AWG used as wavelength optical multiplexer. Since the AWG is able to multiplex N wavelengths (channels) into a single fiber, it increases the number of users up to 40 with less insertion losses.

In a basic setup each user is connected to an output port of the AWG. Each one communicates using a particular wavelength. However, because of the periodicity of the AWG, lower and upper spectrum bands are also multiplexed. Thus, each user is actually able to communicate using more than one wavelength: the wavelength within the spectrum band and the corresponding periodic wavelengths. This allows to address multiple receivers from a single emitter. In this case, care must be taken in that the passband of the OADM coincides with a band of the AWG. The use of several bands in an AWG is a common solution in commercial optical networks for instance to separate downstream from upstream signals~\cite{Park_04}.

Although all the examples comply with the 20 dB loss budget of current commercial QKD systems, a better tolerance to losses and higher bit rates are expected for the next generation of QKD systems~\cite{Stucki_09, Inoue_02, Dixon_08, Namekata_11}. This will allow to put QKD devices in farther away locations and increase the number of disjoint paths, limited to two in the prototypes, thus improving the throughput and security of the network. Moreover, we limit ourselves to the PON realm in order to not have to contend with any active switching elements. If these are included in the network design, they open new possibilities at the expense of having to deal with routing algorithms that might include other security threats.

A rough estimate of the cost, as a function of the number of components, can be used to briefly compare the three proposed prototypes. On one hand, the third prototype is the one with fewer main network components per branch: an optical divisor (AWG), a DWDM OADM connecting the receivers, and a splitter to connect the access and backbone network. On the other, the second prototype requires and additional component since the access network is connected to the backbone using a circulator and a CWDM OADM instead of the splitter. Thus, the third prototype is slightly better than the second relative to the number of components. Finally, the first prototype is clearly a more expensive scenario with five main network components per branch.

\section{Conclusions}
\label{sec:conclusions}

In this paper we introduce weakly trusted repeaters for quantum key distribution networks. The objective is to overcome the distance limitation of QKD with current technology, while improving on the traditional trusted repeater model used in these cases. We formalize these WTR using a network coding approach and construct information theoretically secure scenarios, assumed given specific trust structures. In particular, that there is at least one non-malicious path among the several disjoint paths used. Securing networks with WTR reduces the strong full trust dependence that is assumed for traditional trusted repeaters which, in turn, improves the security of applications and services relying on them. WTR can directly found its niche in organizations with private networks and high security needs, e.g., telecom companies, banks, military institutions or government agencies. In these cases, although nodes belong to the organization, they cannot be fully trusted (possible eavesdropping in the future). WTR can also be useful in networks infrastructure where each path belongs to a different owner. In contrast to the private network situation, here all paths are initially weakly trusted.

Moreover, we have also shown practical scenarios based on optical networks, and detailed implementations with standard components of typical telecom networks. In these networks, WTR can be used along with QKD to exchange secure keys between two users. In contrast with existing proposals, the structure of WTR is more simple and homogeneous, hence facilitating its industrialization. These scenarios can be easily extended to exchange secret keys among a higher number of users and through more intermediate nodes. Considering a higher number of disjoint paths would certainly bring more flexibility to the trust structures. This approach would, as well, open the door to other important topics such as the robustness and resiliency of these networks.

\section*{Acknowledgment}

This work has been partially supported by the project Quantum Information Technologies Madrid\footnote{http://www.quitemad.org} (QUITEMAD, Id. P2009/ESP-1594), funded by \textit{Comunidad Aut\'{o}noma de Madrid} and, Hybrid Quantum Networks (HyQuNet, Id. TEC2012-35673), funded by \textit{Ministerio de Econom\'{i}a y Competitividad}, Spain.

\bibliographystyle{IEEEtran}

\begin{thebibliography}{99}

\bibitem{Senior_08}
J.~M. Senior and Y.~Jamro, \emph{{Optical Fiber Communications: Principles and
  Practice}}. Financial Times/Prentice Hall, 2008.

\bibitem{Shaneman_04}
K.~Shaneman and S.~Gray, ``Optical network security: technical analysis of
  fiber tapping mechanisms and methods for detection prevention,'' in
  \emph{MILCOM 2004, IEEE Military Communications Conference}, vol.~2, 2004,
  pp. 711--716.

\bibitem{Fok_11}
M.~P. Fok, Z.~Wang, Y.~Deng, and P.~R. Prucnal, ``{Optical Layer Security in
  Fiber-Optic Networks},'' \emph{IEEE Trans. Inf. Forensic Secur.}, vol.~6,
  no.~3, pp. 725--736, 2011.

\bibitem{Gutierrez_07}
D.~Gutierrez, J.~Cho, and L.~G. Kazovsky, ``{TDM-PON Security Issues: Upstream
  Encryption is Needed},'' in \emph{OFC/NFOEC 2007, Conference on Optical Fiber
  Communication and the National Fiber Optic Engineers Conference}, 2007, pp.
  1--3.

\bibitem{Kazovsky_11}
L.~G. Kazovsky, S.-W. Wong, V.~Gudla, P.~T. Afshar, S.-H. Yen, S.~Yamashita,
  and Y.~Yan, ``{Challenges in next-generation optical access networks:
  addressing reach extension and security weaknesses},'' \emph{IET
  Optoelectron.}, vol.~5, no.~4, pp. 133--143, 2011.

\bibitem{Atkins_95}
D.~Atkins, M.~Graff, A.~K. Lenstra, and P.~C. Leyland, ``{The Magic Words are
  Squeamish Ossifrage},'' in \emph{Proceedings of the 4th International
  Conference on the Theory and Applications of Cryptology: Advances in
  Cryptology}, 1995, pp. 263--277.

\bibitem{RSA}
{Key Size Recommendations by RSA Laboratories}. [Online]. Available:
  \emph{http://www.rsa.com/rsalabs/node.asp?id=2004}

\bibitem{Barker_06}
E.~Barker, W.~Barker, W.~Burr, W.~Polk, and M.~Smid, ``{Recommendation for key
  management - part 1: General (revised)},'' in \emph{NIST Special
  Publication}, 2006.

\bibitem{Gisin_02}
N.~Gisin, G.~Ribordy, W.~Tittel, and H.~Zbinden, ``{Quantum Cryptography},''
  \emph{Rev. Mod. Phys.}, vol.~74, no.~1, pp. 145--195, 2002.

\bibitem{Stucki_11}
D.~Stucki, M.~Legr\'{e}, F.~Buntschu, B.~Clausen, N.~Felber, N.~Gisin,
  L.~Henzen, P.~Junod, G.~Litzistorf, P.~Monbaron, L.~Monat, J.-B. Page,
  D.~Perroud, G.~Ribordy, A.~Rochas, S.~Robyr, J.~Tavares, R.~Thew,
  P.~Trinkler, S.~Ventura, R.~Voirol, N.~Walenta, and H.~Zbinden, ``{Long-term
  performance of the SwissQuantum quantum key distribution network in a field
  environment},'' \emph{New J. Phys.}, vol.~13, no.~12, p. 123001, 2011.

\bibitem{Jouguet_12}
P.~Jouguet, S.~Kunz-Jacques, T.~Debuisschert, S.~Fossier, E.~Diamanti,
  R.~All\'{e}aume, R.~Tualle-Brouri, P.~Grangier, A.~Leverrier, P.~Pache, and
  P.~Painchault, ``Field test of classical symmetric encryption with continuous
  variables quantum key distribution,'' \emph{Opt. Express}, vol.~20, no.~13,
  pp. 14\,030--14\,041, 2012.

\bibitem{Wootters_82}
W.~K. Wootters and W.~H. Zurek, ``{A single quantum cannot be cloned},''
  \emph{Nature}, vol. 299, no. 5886, pp. 802--803, 1982.

\bibitem{Townsend_94}
P.~D. Townsend, S.~J.~D. Phoenix, K.~J. Blow, and S.~M. Barnett, ``Design of
  quantum cryptography systems for passive optical networks,'' \emph{Electron.
  Lett.}, vol.~30, no.~22, pp. 1875--1877, 1994.

\bibitem{Toliver_03}
P.~Toliver, R.~J. Runser, T.~E. Chapuran, J.~L. Jackel, T.~C. Banwell, M.~S.
  Goodman, R.~J. Hughes, C.~G. Peterson, D.~Derkacs, J.~E. Nordholt, L.~Mercer,
  S.~McNown, A.~Goldman, and J.~Blake, ``Experimental investigation of quantum
  key distribution through transparent optical switch elements,'' \emph{IEEE
  Photonics Technol. Lett.}, vol.~15, no.~11, pp. 1669--1671, 2003.

\bibitem{Lancho_09}
D.~Lancho, J.~Mart\'{i}nez, D.~Elkouss, M.~Soto, and V.~Mart\'{i}n, ``{QKD in
  Standard Optical Telecommunications Networks},'' in \emph{1st International
  Conference on Quantum Communication and Quantum Networking}, vol.~36, 2010,
  pp. 142--149.

\bibitem{Maeda_09}
W.~Maeda, A.~Tanaka, S.~Takahashi, A.~Tajima, and A.~Tomita, ``{Technologies
  for Quantum Key Distribution Networks Integrated With Optical Communication
  Networks},'' \emph{IEEE J. Sel. Top. Quantum Electron.}, vol.~15, no.~6, pp.
  1591--1601, 2009.

\bibitem{Chapuran_09}
T.~E. Chapuran, P.~Toliver, N.~A. Peters, J.~Jackel, M.~S. Goodman, R.~J.
  Runser, S.~R. McNown, N.~Dallmann, R.~J. Hughes, K.~P. McCabe, J.~E.
  Nordholt, C.~G. Peterson, K.~T. Tyagi, L.~Mercer, and H.~Dardy, ``Optical
  networking for quantum key distribution and quantum communications,''
  \emph{New J. Phys.}, vol.~11, no.~10, p. 105001, 2009.

\bibitem{BingQi_10}
B.~Qi, W.~Zhu, L.~Qian, and H.-K. Lo, ``Feasibility of quantum key distribution
  through a dense wavelength division multiplexing network,'' \emph{New J.
  Phys.}, vol.~12, no.~10, p. 103042, 2010.

\bibitem{Kitayama_11}
K.-I. Kitayama, M.~Sasaki, S.~Araki, M.~Tsubokawa, A.~Tomita, K.~Inoue,
  K.~Harasawa, Y.~Nagasako, and A.~Takada, ``{Security in Photonic Networks:
  Threats and Security Enhancement},'' \emph{J. Lightwave Technol.}, vol.~29,
  no.~21, pp. 3210--3222, 2011.

\bibitem{Wang_12}
S.~Wang, W.~Chen, J.~Guo, Z.~Yin, H.~Li, Z.~Zhou, G.~Guo, and Z.~Han, ``{2 GHz
  clock quantum key distribution over 260 km of standard telecom fiber},''
  \emph{Opt. Lett.}, vol.~37, no.~6, pp. 1008--1010, 2012.

\bibitem{Elliot_02}
C.~Elliot, ``Building the quantum network,'' \emph{New J. Phys.}, vol.~4,
  no.~1, pp. 46.1--46.12, 2002.

\bibitem{Peev_09}
M.~Peev, C.~Pacher, R.~All\'{e}aume, C.~Barreiro, J.~Bouda, W.~Boxleitner,
  T.~Debuisschert, E.~Diamanti, M.~Dianati, J.~F. Dynes, S.~Fasel, S.~Fossier,
  M.~F\"{u}rst, J.-D. Gautier, O.~Gay, N.~Gisin, P.~Grangier, A.~Happe,
  Y.~Hasani, M.~Hentschel, H.~H\"{u}bel, G.~Humer, T.~L\"{a}nger, M.~Legr\'{e},
  R.~Lieger, J.~Lodewyck, T.~Lor\"{u}nser, N.~L\"{u}tkenhaus, A.~Marhold,
  T.~Matyus, O.~Maurhart, L.~Monat, S.~Nauerth, J.-B. Page, A.~Poppe,
  E.~Querasser, G.~Ribordy, S.~Robyr, L.~Salvail, A.~W. Sharpe, A.~J. Shields,
  D.~Stucki, M.~Suda, C.~Tamas, T.~Themel, R.~T. Thew, Y.~Thoma, A.~Treiber,
  P.~Trinkler, R.~Tualle-Brouri, F.~Vannel, N.~Walenta, H.~Weier,
  H.~Weinfurter, I.~Wimberger, Z.~L. Yuan, H.~Zbinden, and A.~Zeilinger, ``{The
  SECOQC quantum key distribution network in Vienna},'' \emph{New J. Phys.},
  vol.~11, no.~7, p. 075001, 2009.

\bibitem{Sasaki_11}
M.~Sasaki, M.~Fujiwara, H.~Ishizuka, W.~Klaus, K.~Wakui, M.~Takeoka, S.~Miki,
  T.~Yamashita, Z.~Wang, A.~Tanaka, K.~Yoshino, Y.~Nambu, S.~Takahashi,
  A.~Tajima, A.~Tomita, T.~Domeki, T.~Hasegawa, Y.~Sakai, H.~Kobayashi,
  T.~Asai, K.~Shimizu, T.~Tokura, T.~Tsurumaru, M.~Matsui, T.~Honjo, K.~Tamaki,
  H.~Takesue, Y.~Tokura, J.~Dynes, A.~Dixon, A.~Sharpe, Z.~Yuan, A.~Shields,
  S.~Uchikoga, M.~Legr\'{e}, S.~Robyr, P.~Trinkler, L.~Monat, J.~Page,
  G.~Ribordy, A.~Poppe, A.~Allacher, O.~Maurhart, T.~L\"{a}nger, M.~Peev, and
  A.~Zeilinger, ``{Field test of quantum key distribution in the Tokyo QKD
  Network},'' \emph{Opt. Express}, vol.~19, no.~11, pp. 10\,387--10\,409, 2011.

\bibitem{Briegel_98}
H.-J. Briegel, W.~D\"{u}r, J.~I. Cirac, and P.~Zoller, ``{Quantum Repeaters:
  The Role of Imperfect Local Operations in Quantum Communication},''
  \emph{Phys. Rev. Lett.}, vol.~81, no.~26, pp. 5932--5935, 1998.

\bibitem{Alleaume_09}
R.~All\'{e}aume, F.~Roueff, E.~Diamanti, and N.~L\"{u}tkenhaus, ``{Topological
  optimization of quantum key distribution networks},'' \emph{New J. Phys.},
  vol.~11, no.~7, p. 075002, 2009.

\bibitem{NIST_01}
NIST, ``Security requirements for cryptographic modules,'' \emph{FIPS PUB}, no.
  140-2, 2001.

\bibitem{CommonCriteria}
{Common Criteria}. [Online]. Available:
  \emph{http://www.commoncriteriaportal.org}

\bibitem{Ahlswede_00}
R.~Ahlswede, N.~Cai, S.-Y. Li, and R.~Yeung, ``{Network information flow},''
  \emph{IEEE Trans. Inf. Theory}, vol.~46, no.~4, pp. 1204--1216, 2000.

\bibitem{Cai_11b}
N.~Cai and T.~Chan, ``{Theory of Secure Network Coding},'' \emph{Proceedings of
  the IEEE}, vol.~99, no.~3, pp. 421--437, 2011.

\bibitem{Manley_08}
E.~D. Manley, J.~S. Deogun, L.~Xu, and D.~R. Alexander, ``{Network Coding for
  WDM All-Optical Multicast},'' University of Nebraska, Lincoln, Tech. Rep.,
  2008.

\bibitem{Kamal_10}
A.~E. Kamal, A.~Ramamoorthy, L.~Long, and L.~Shizheng, ``{Overlay Protection
  Against Link Failures Using Network Coding},'' \emph{IEEE-ACM Trans. Netw.},
  vol.~19, no.~4, pp. 1071--1084, 2011.

\bibitem{Manley_10}
E.~D. Manley, J.~S. Deogun, L.~Xu, and D.~R. Alexander, ``{All-Optical Network
  Coding},'' \emph{J. Opt. Commun. Netw.}, vol.~2, no.~4, pp. 175--191, 2010.

\bibitem{Belzner_09}
M.~Belzner and H.~Haunstein, ``Network coding in passive optical networks,'' in
  \emph{ECOC 2009, 35th European Conference on Optical Communication}, 2009,
  pp. 1--2.

\bibitem{Miller_10}
K.~Miller, T.~Biermann, H.~Woesner, and H.~Karl, ``{Network Coding in Passive
  Optical Networks},'' in \emph{NetCod 2010, IEEE International Symposium on
  Network Coding}, 2010, pp. 1--6.

\bibitem{Fouli_11}
K.~Fouli, M.~Maier, and M.~Medard, ``Network coding in next-generation passive
  optical networks,'' \emph{IEEE Commun. Mag.}, vol.~49, no.~9, pp. 38--46,
  2011.

\bibitem{Cover_91}
T.~M. Cover and J.~A. Thomas, \emph{{Elements of Information Theory}}.\hskip
  1em plus 0.5em minus 0.4em\relax Wiley-Interscience, 1991.

\bibitem{Chan_08}
T.~Chan and A.~Grant, ``{Capacity Bounds for Secure Network Coding},'' in
  \emph{Proceedings of Comm. Theory Workshop}, 2008, pp. 94--100.

\bibitem{Dolev_93}
D.~Dolev, C.~Dwork, O.~Waarts, and M.~Yung, ``Perfectly secure message
  transmission,'' \emph{J. ACM}, vol.~40, pp. 17--47, 1993.

\bibitem{Jaggi_08}
S.~Jaggi, M.~Langberg, S.~Katti, T.~Ho, D.~Katabi, M.~Medard, and M.~Effros,
  ``{Resilient Network Coding in the Presence of Byzantine Adversaries},''
  \emph{IEEE Trans. Inf. Theory}, vol.~54, no.~6, pp. 2596--2603, 2008.

\bibitem{Salvail_10}
L.~Salvail, M.~Peev, E.~Diamanti, R.~All\'{e}aume, N.~L\"{u}tkenhaus, and
  T.~L\"{a}nger, ``{Security of trusted repeater quantum key distribution
  networks},'' \emph{Journal of Computer Security}, vol.~18, no.~1, pp. 61--87,
  2010.

\bibitem{Gobby_04}
C.~Gobby, Z.~L. Yuan, and A.~J. Shields, ``{Quantum key distribution over 122
  km of standard telecom fiber},'' \emph{Appl. Phys. Lett.}, vol.~84, no.~19,
  p. 3762, 2004.

\bibitem{idQuantique}
{ID Quantique SA}. [Online]. Available: \emph{http://www.idquantique.com}

\bibitem{Jouguet_12b}
P.~Jouguet, S.~Kunz-Jacques, A.~Leverrier, P.~Grangier, and E.~Diamanti,
  ``Experimental demonstration of continuous-variable quantum key distribution
  over 80 km of standard telecom fiber,'' in \emph{QCRYPT 2012, 2nd Annual
  Conference on Quantum Cryptography}, 2012.

\bibitem{Alba_12}
A.~Ruiz-Alba, J.~Mora, W.~Amava, A.~Mart\'{i}nez, V.~Garc\'{i}a-Mu\~{n}oz,
  D.~Calvo, and J.~Capmany, ``{Microwave Photonics Parallel Quantum Key
  Distribution},'' \emph{IEEE Photonics J.}, vol.~4, no.~3, pp. 931--942, 2012.

\bibitem{Ma_05}
X.~Ma, H.-K. Lo, Y.~Zhao, and B.~Qi, ``{Practical decoy state for quantum key
  distribution},'' \emph{Phys. Rev. A}, vol.~72, no.~1, p. 012326, 2005.

\bibitem{Masanes_11}
L.~Masanes, S.~Pironio, and A.~Ac\'{i}n, ``Secure device-independent quantum
  key distribution with causally independent measurement devices,'' \emph{Nat.
  Commun.}, vol.~2, p. 238, 2011.

\bibitem{Ramaswami_09}
R.~Ramaswami, K.~Sivarajan, and G.~Sasaki, \emph{{Optical Networks: A Practical
  Perspective}}.\hskip 1em plus 0.5em minus 0.4em\relax Morgan Kaufmann
  Publishers Inc., 2009.

\bibitem{Choi_10}
I.~Choi, R.~J. Young, and P.~D. Townsend, ``{Quantum key distribution on a
  10Gb/s WDM-PON},'' \emph{Opt. Express}, vol.~18, no.~9, pp. 9600--9612, 2010.

\bibitem{Eraerds_10}
P.~Eraerds, N.~Walenta, M.~Legr\'{e}, N.~Gisin, and H.~Zbinden, ``{Quantum key
  distribution and 1 Gbps data encryption over a single fibre},'' \emph{New J.
  Phys.}, vol.~12, no.~6, p. 063027, 2010.

\bibitem{Park_04}
S.-J. Park, C.-H. Lee, K.-T. Jeong, H.-J. Park, J.-G. Ahn, and K.-H. Song,
  ``Fiber-to-the-home services based on wavelength-division-multiplexing
  passive optical network,'' \emph{J. Lightwave Technol.}, vol.~22, no.~11, pp.
  2582--2591, 2004.

\bibitem{Stucki_09}
D.~Stucki, N.~Walenta, F.~Vannel, R.~T. Thew, N.~Gisin, H.~Zbinden, S.~Gray,
  C.~R. Towery, and S.~Ten, ``High rate, long-distance quantum key distribution
  over 250 km of ultra low loss fibres,'' \emph{New J. Phys.}, vol.~11, no.~7,
  p. 075003, 2009.

\bibitem{Inoue_02}
K.~Inoue, E.~Waks, and Y.~Yamamoto, ``Differential phase shift quantum key
  distribution,'' \emph{Phys. Rev. Lett.}, vol.~89, no.~3, p. 037902, 2002.

\bibitem{Dixon_08}
A.~R. Dixon, Z.~L. Yuan, J.~F. Dynes, A.~W. Sharpe, and A.~J. Shields,
  ``{Gigahertz decoy quantum key distribution with 1 Mbit/s secure key rate},''
  \emph{Opt. Express}, vol.~16, no.~23, pp. 18\,790--18\,979, 2008.

\bibitem{Namekata_11}
N.~Namekata, H.~Takesue, T.~Honjo, Y.~Tokura, and S.~Inoue, ``{High-rate
  quantum key distribution over 100 km using ultra-low-noise, 2-GHz
  sinusoidally gated InGaAs/InP avalanche photodiodes},'' \emph{Opt. Express},
  vol.~19, no.~11, pp. 10\,632--10\,639, 2011.

\end{thebibliography}

\end{document}